\RequirePackage{fix-cm}
\documentclass[twocolumn,epjc3]{svjour3}  
\smartqed  
\RequirePackage{graphicx}
\RequirePackage{amssymb}
\usepackage{gensymb}
\usepackage{dblfloatfix} 
\usepackage[caption=false]{subfig}
\usepackage{changes}
\usepackage[numbers]{natbib}
\usepackage[switch]{lineno}
\usepackage{overpic}
\usepackage{orcidlink}
\usepackage{array}
\newcolumntype{P}[1]{>{\centering\arraybackslash}p{#1}}
\journalname{Eur. Phys. J. C}
\begin{document}

\title{Muons as a tool for background rejection in imaging atmospheric telescope arrays}

\titlerunning{Muons for background rejection in IACTs}  

\author{L. Olivera-Nieto \orcidlink{0000-0002-9105-0518}\thanksref{e1,addr1}
        \and
        A. M. W. Mitchell \orcidlink{0000-0003-3631-5648}\thanksref{addr2, addr3} 
        \and 
        K. Bernl\"{o}hr \orcidlink{0000-0001-8065-3252}\thanksref{addr1}
        \and
         J.~A.~Hinton~\orcidlink{0000-0002-1031-7760}\thanksref{addr1}\hspace{-0.4 cm} }

\thankstext{e1}{e-mail: Laura.Olivera-Nieto@mpi-hd.mpg.de}


\institute{Max-Planck-Institut für Kernphysik, P.O. Box 103980, D 69029, Heidelberg, Germany \label{addr1}
           \and
           Department of Physics, ETH Zurich,  CH-8093 Zurich, Switzerland \label{addr2}
           \and
           Friedrich-Alexander-Universit\"at Erlangen-N\"urnberg, Erlangen Centre for Astroparticle Physics, 91058 Erlangen, Germany \label{addr3}
}

\date{Received: date / Accepted: date}

\maketitle

\begin{abstract}
The presence of muons in air-showers initiated by cosmic ray protons and nuclei is well established as a powerful tool to separate such showers from those initiated by gamma rays. However, so far this approach has been fully exploited only for ground level particle detecting arrays. We explore the feasibility of using Cherenkov light from muons as a background rejection tool for imaging atmospheric Cherenkov telescope arrays at the highest energies. We adopt an analytical model of the Cherenkov light from individual muons to allow rapid simulation of a large number of showers in a hybrid mode. This allows us to explore the very high background rejection power regime at acceptable cost in terms of computing time.
We show that for very large ($\gtrsim$20~m mirror diameter) telescopes, efficient identification of muon light can potentially lead to background rejection levels up to 10$^{-5}$ whilst retaining high efficiency for gamma rays. While many challenges remain in the effective exploitation of the muon Cherenkov light in the data analysis for imaging Cherenkov telescope arrays, our study indicates that for arrays containing at least one large telescope, this is a very worthwhile endeavor.
\end{abstract}

\section{Introduction}
\label{sec:intro}
\begin{figure*}
    \centering
    \begin{overpic}[width=1.9\columnwidth]{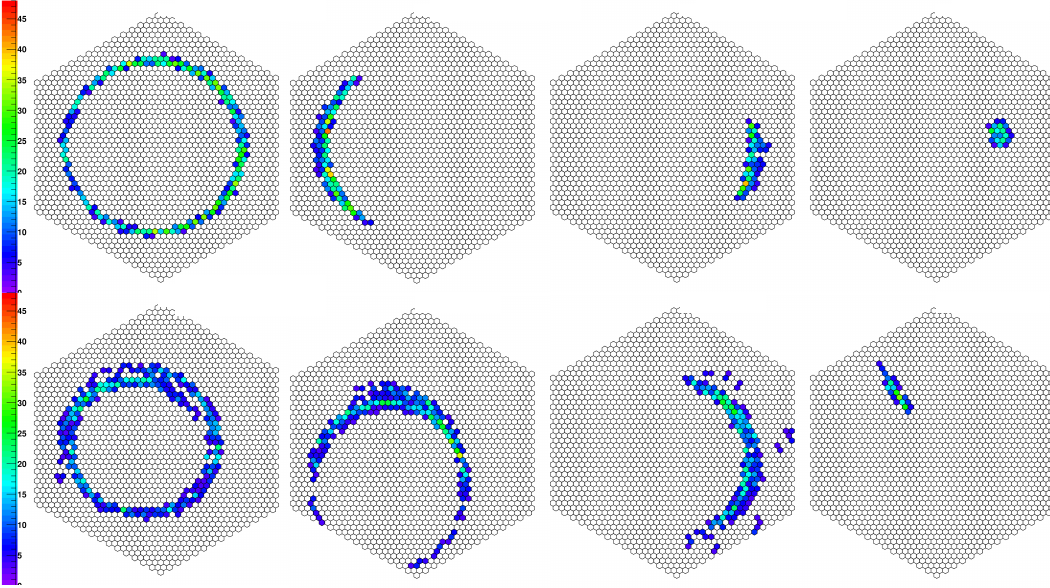}
    \put(3,54){E$>$20~GeV}
    \put(3,26){E$<$10~GeV}
    \put(21,52){$\sim$0~m}
    \put(45,52){$\sim$15~m}
    \put(70,52){$\sim$30~m}
    \put(95,52){$\sim$100~m}
    
    \put(21,24){$\sim$0~m}
    \put(45,24){$\sim$15~m}
    \put(70,24){$\sim$30~m}
    \put(95,24){$\sim$100~m}
    \end{overpic}
    \caption{Example simulated muon images in a 28~m telescope at different energies and impact distances. All simulated muons are produced at 11~km above sea level and are observed at 1835~m with a zenith angle of 20\degree.}
    \label{fig:images}
\end{figure*}
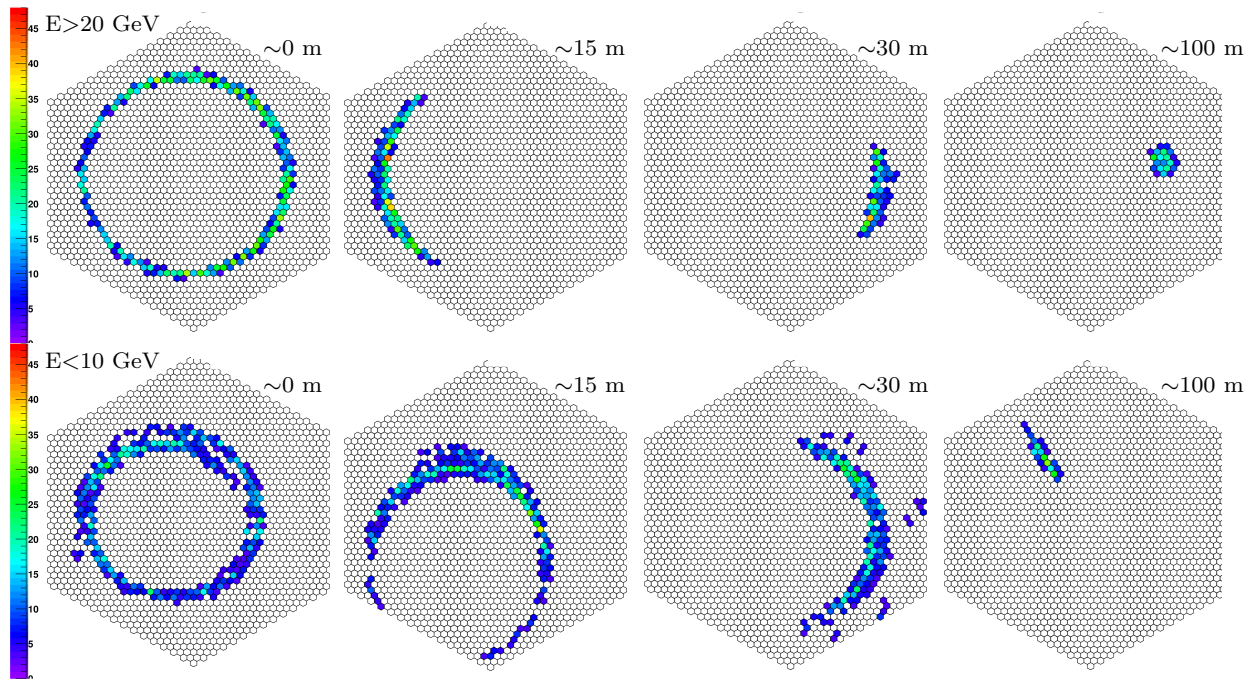

The hadronic cascades associated to charged cosmic ray primary particles typically produce large numbers of muons, primarily from the decay of charged pions. The potential to use these muons to discriminate between hadronic and electromagnetic cascades and hence do gamma-ray astronomy, has long been recognised (see for example~\cite{Gaisser}). 

The performance of the LHAASO array~\cite{lhaaso,lhaaso-crab}, demonstrates the power of this approach for very-high and ultra-high energy gamma-ray astronomy (from tens to hundreds of TeV). An important factor contributing to the success of the 
LHAASO array is the very large total area of muon detectors, a factor of  nearly 20 larger than the CASA-MIA~\cite{casa-mia} array, resulting in a more than hundredfold improvement in background rejection. 

Ground level muons become a useful separation tool for showers above $\sim$1~TeV at high altitude~\cite{harmpaper}. However, excellent hadron rejection power, that is, over a factor $10^{4}$ reduction, is possible only at tens of~TeV \cite{lhaaso-crab}.

Imaging Atmospheric Cherenkov Telescope (IACT) arrays have superior rejection power to other ground-based arrays in the domain around 1~TeV, exploiting primarily the differences in shower width and substructure between electromagnetic and hadronic showers in this energy range~\cite{DanStefan,stefanBDT,magicBDT}. However, IACTs have not so far demonstrated excellent rejection power at tens of TeV. Traditional separation approaches are limited by the fact that, beyond small impact distances, large events are typically not fully contained by the camera image, which significantly affects the estimation of the necessary shower parameters. The current background rejection power attained by the traditional separation methods at energies above a few tens of TeVs reaches levels between $10^{-2}$ and $10^{-3}$~\cite{DanStefan, stefanBDT,berge}.
This loss of performance at high energies can also be seen, albeit indirectly, in the expected background rate after background rejection cuts for CTA\footnote{\url{https://www.cta-observatory.org/science/cta-performance/\#1472563453568-c1970a6e-2c0f}}. 
The surviving background rate falls by a factor 500 between 0.1 and 1 TeV but less than a factor of 10 between 10 TeV and a 100 TeV - while the proton flux falls by a factor of 50 per decade. 

The ring-like images produced when ground-level muons pass through IACTs have long been used as a means of calibration~\cite{vacanti-et-all}, see~\cite{ctamuoncalibration} for a recent review.
Recently,~\cite{alisonmuons} suggested that the identification of a much higher fraction of muons produced in extensive air showers is possible with IACTs.

Large telescopes such as the central telescope of H.E.S.S.~\cite{hess2} and the Large-Sized Telescopes (LSTs) of CTA~\cite{cta} enable the detection of individual muons out to large impact distance. This has traditionally been seen as a problem 
due to their apparent similarity to gamma-like events~\cite{2007APh....28...72M}, but can also be seen as an opportunity for improvement of the background rejection power at the highest energies if characteristic differences of muons to gamma rays are identifiable.

In this paper we explore the potential for muon measurement with IACTs as a tool for background rejection by characterizing the number of muons that are detectable by a large Cherenkov telescope in proton- and gamma-initiated showers of different energies. In our simulations we adopt a hybrid approach to allow exploration of the very high background rejection power regime at acceptable cost in terms of computing time. This approach is introduced and motivated in Section~\ref{sec:muons}. Section~\ref{sec:showers} discusses the muon content of showers from the perspective of air-Cherenkov detection and details the criteria used to label muons as detectable by a telescope. Section~\ref{sec:results} presents the result of said criteria applied to showers initiated by both protons and gamma rays, both for large- and medium-dish telescopes. Finally, Section~\ref{sec:discussion}  discusses the implications and potential for muons as a means to improving IACT background rejection.

\begin{table*}[t]
\label{tab:compare}  
\centering
\begin{tabular}{p{4cm} P{6cm} P{6cm}}
\noalign{\smallskip}\hline\noalign{\smallskip}
     & \centering CORSIKA + \textit{sim\_telarray}  &  Simplified  Muon Model \\
\noalign{\smallskip}\hline\noalign{\smallskip}
Cherenkov Light Production & \multicolumn{2}{c}{Ignoring wavelength dependence of refractive index\footnotemark} \\
Atmospheric Absorption & \multicolumn{2}{c}{Wavelength-dependent tabulated atmospheric absorption~\cite{atmosphere}} \\
Atmosphere characterization &  \multicolumn{2}{c}{Tabulated atmospheric profiles at H.E.S.S. location~\cite{sim-telarray}} \\
Muon scattering & Full treatment & Ignored \\
Muon bremstrahlung & EGS4~\cite{EGS4} & Ignored \\
Ionization losses & Bethe-Bloch formula & 2 MeV per g/cm$^2$\\
Telescope response & Full treatment  & Full treatment \\
Ray-tracing & Full treatment  & Simplified \\
Camera trigger & \multicolumn{2}{c}{Patch of 9 neighboring pixels with total intensity above 68 p.e} \\
Pixel shape & Realistic, hexagonal & Simplified, square \\
Night sky background & Optional, ignored here & Ignored \\
Bending in geomagnetic field & Included & Ignored \\
\noalign{\smallskip}\hline
\end{tabular}
\caption{Comparison of muon treatment between the CORSIKA+\textit{sim\_telarray} approach and the simplified muon model.}
\end{table*}

\section{Cherenkov light from muons}
\label{sec:muons}

The properties of the Cherenkov emission of a single atmospheric muon are very well determined and straight-forward to calculate in comparison to air-showers in general. The suppressed bremsstrahlung cross-section of muons with respect to electrons allows a majority of muons to reach ground-level with only ionisation losses. Similarly, the reduced multiple scattering of muons with respect to electrons means that the assumption of a linear trajectory is reasonable in most cases. For these reasons the simulation of muons in full detail may not be necessary to capture the essential characteristics of Cherenkov light from muons in showers. We implement a simplified muon model (SMM) which, starting from basic muon properties, approximates the Cherenkov light production and telescope simulation with an analytical treatment, described below.  Table~\ref{tab:compare} compares the SMM approach to the combination of the CORSIKA~\cite{corsika} package for shower and Cherenkov light simulation and the \textit{sim\_telarray}~\cite{sim-telarray}  package for the telescope response and camera simulation. In order to verify the predictions of the SMM, we produced a small set of full CORSIKA+\textit{sim\_telarray} muon simulations with energies between 5 and 100~GeV for different starting heights. Note that the Cherenkov threshold for muons at 1835~m above sea level is slightly above 5~GeV. 

The key parameters affecting the Cherenkov image properties of individual muons are the initial energy and production height in the atmosphere. Muons that reach ground level and land close to, or intersect, a telescope dish, produce a ring-shaped image in the telescope camera, with a full circle for muons hitting the dish and reduced sections of arc as the impact distance becomes larger. The surface brightness of the images, however, remains mostly constant, which allows muons to trigger out to large impact distances, even when the ring section captured by the camera is small enough to no longer resemble an arc, but rather a small cluster of pixels.
Figure~\ref{fig:images} illustrates the evolution of muon image properties with impact distance, as imaged in the 28~m telescope of H.E.S.S., based on CORSIKA and \textit{sim\_telarray} simulations.

The first step to determine whether an incoming muon can be detected by a telescope located at a certain distance from its ground impact point, is to compute the amount of Cherenkov photons collected by the telescope camera as a function of said distance. This distribution is calculated assuming a straight trajectory of the muon through the atmosphere from the production height h$_{\mathrm{prod}}$. The atmospheric density $\rho(h)$ and refractive index $n(h)$ profiles are described by the same model used by \textit{sim\_telarray} at the H.E.S.S. location~\cite{sim-telarray}. The wavelength dependence of the refractive index is ignored. For a muon with incoming zenith angle $\theta_z$, the actual path through the atmosphere is then described by $l = h/\cos\theta_z$ for $h \in [h_{\mathrm{ground}}, h_{\mathrm{prod}}]$, where $h_{\mathrm{ground}}$ is taken to be 1835~m above sea level. The emitted photons are subject to wavelength-dependent atmospheric absorption A$(\lambda,h)$, which is integrated along the photon path, assumed here for simplicity to be the same as the muon path starting at the point where the photon is produced. 
The photons produced at a height $h$ then arrive at the ground at a distance $R(h)$ from the point where the muon hits the ground 
\begin{equation}
\label{eq:impact}
    R(h) = (h-h_{\mathrm{ground}})\cdot\frac{\sin\theta_c}{\cos \theta_z},
\end{equation}
where $\theta_c = \arccos((n\cdot \beta)^{-1})$ is the Cherenkov angle of a muon traveling with velocity $\beta= v/c$ in a medium with refractive index $n$. Note that $\theta_c$ varies with height, because of energy losses and the refractive index profile.

The number of Cherenkov photons $N_{\gamma}$ initially produced by the muon per path length $dl$ between wavelengths $\lambda_1$ and $\lambda_2$, is described by 
\begin{equation}
\label{eq:yield}
\frac{dN_{\gamma}}{dl} = 2\pi\alpha \left( 1-(\beta n)^{-2}\right)\int_{\lambda_1}^{\lambda_2} \frac{d\lambda}{\lambda^2},
\end{equation}
where $\alpha$ is the fine-structure constant. This quantity needs to be convolved with the telescope response, which consists of many different elements. 
For this, we used the response of the telescopes in the H.E.S.S. array. The wavelength-dependent telescope response W$_T(\lambda)$ is the combination of the mirror reflectivity, quantum efficiency of the camera and plexiglas transmittance of the camera window. Additionally, there are wavelength-independent corrections for the telescope area projection and the camera and Winston cones shadowing, which combine to a factor $f \sim 0.6$ for a 28~m telescope.
\footnotetext{Note that CORSIKA can generate photons according to a wavelength-dependent index of refraction but it is not used by default for reasons of computing efficiency.} Combining the number of photons initially produced with the telescope response results in the number of Cherenkov photons produced by a muon per unit path length that are detected by the telescope:
\begin{equation}
\label{eq:yield_tel}
\frac{dN_{\gamma, T}}{dl} = 2f\pi\alpha \left( 1-(\beta n)^{-2}\right)\int_{\lambda_1}^{\lambda_2} A(\lambda, h) W_T(\lambda)\frac{d\lambda}{\lambda^2}.
\end{equation}
\begin{figure}[h!]
\subfloat{%
  \includegraphics[clip,width=\columnwidth]{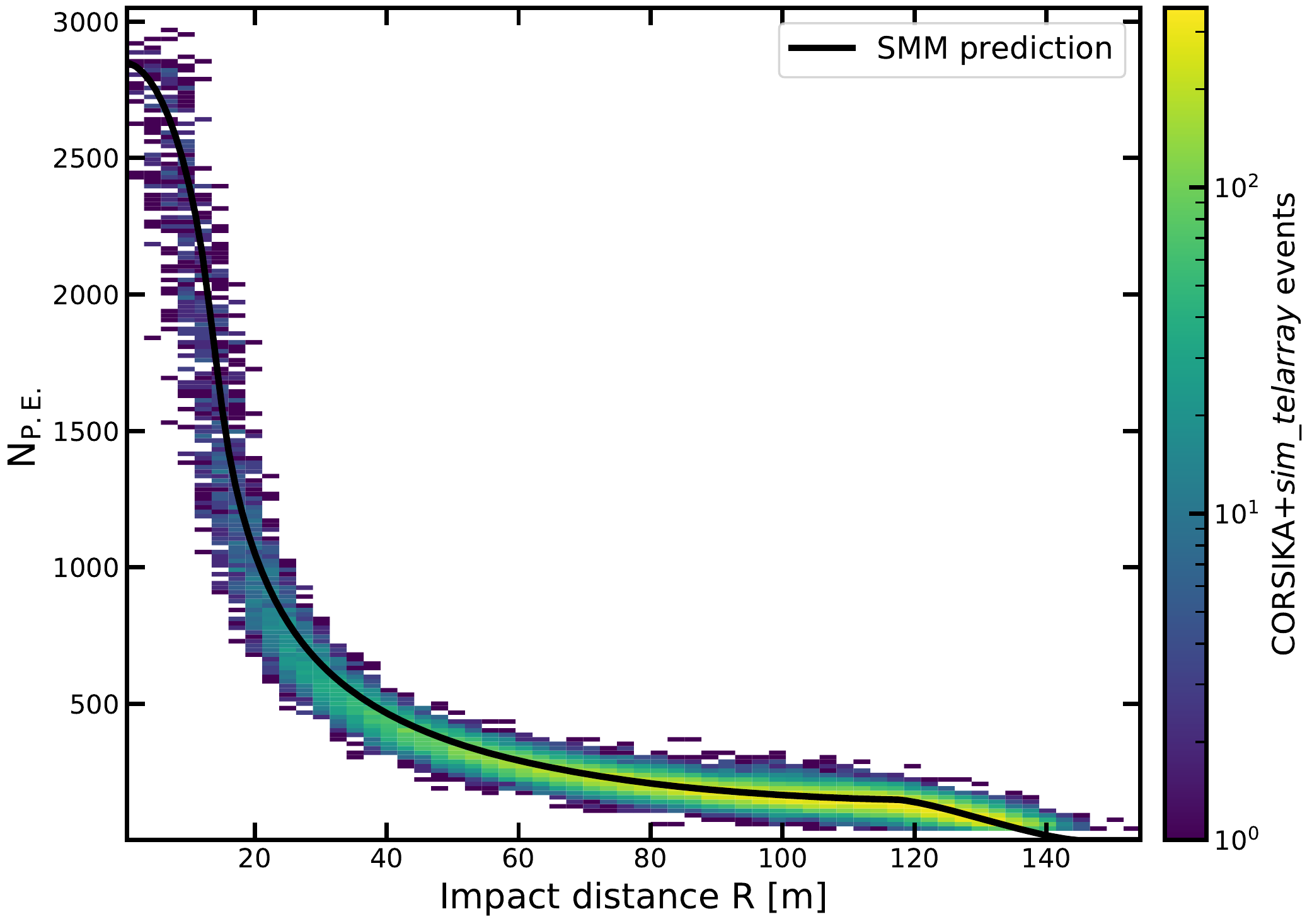}
  }\\
\subfloat{%
  \includegraphics[clip,width=0.95\columnwidth]{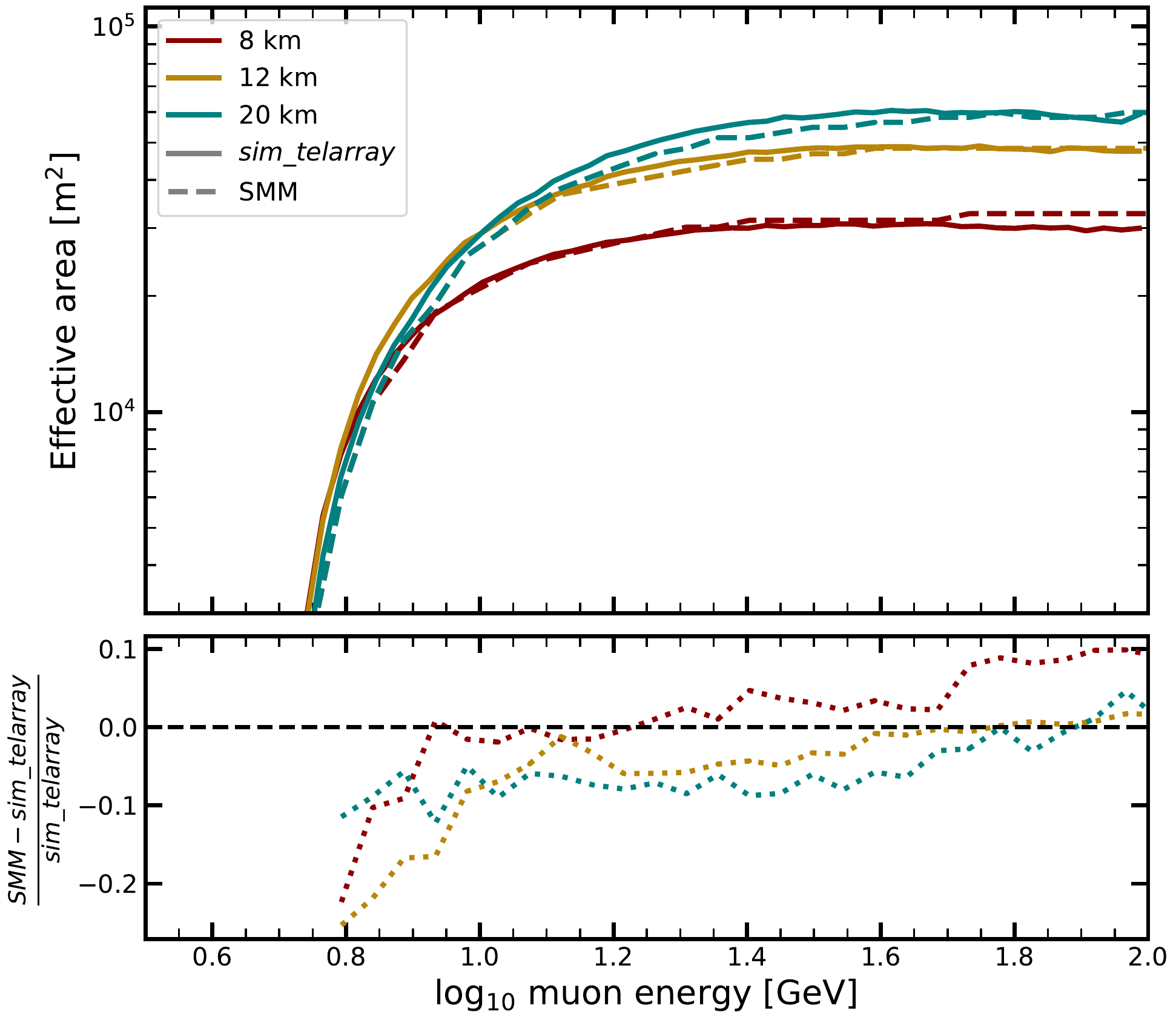}
  }
\caption{\textit{Top}: Comparison of the amount of photoelectrons predicted by the analytical model for a 28~m dish (black line) and the resulting distribution for simulated muon showers for energies above 20~GeV, scattering distance smaller than 5~m and starting height $\sim$ 10~km. 
\textit{Bottom}: Effective area comparison between the simulations and the simplified model for different starting heights. Includes all events, even those that undergo significant scattering and bremsstrahlung. }
\label{fig:toycompare}       
\end{figure}
The photons distribute radially on the ground from the impact position of the muon, which defines the origin of the coordinate system. Using Equations~\ref{eq:impact} and~\ref{eq:yield_tel} we can compute the ground density of detected photons
\begin{equation}
    \label{eq:dens_tel}
    \rho_{\gamma, T}(R) = \frac{dN_{\gamma, T}}{dA} = \frac{dN_{\gamma, T}}{2\pi R dR} = \frac{dN_{\gamma, T}}{dl}\frac{dl}{2\pi R dR},
\end{equation}
where R is the distance to the muon impact position. Placing a test telescope with a circular mirror of diameter D$_T$ in the ground at position $(x = 0, y = R)$, the amount of photoelectrons (P.E.) collected by the dish as a function of $R$ is then given by
\begin{equation}
    \label{eq:pe}       
    \mathrm{N_{P.E.}}(R) = \int_{R-R_T}^{R+R_T}dx\int_{-\alpha}^{\alpha}dy \cdot \rho_{\gamma, T}({\scriptstyle \sqrt{x^2 + y^2}}),
\end{equation}
where $\alpha = \sqrt{R_T^2-(x-R)^2}$.

The number of photoelectrons predicted by the SMM is compared in the top panel of Figure~\ref{fig:toycompare} with that resulting from the CORSIKA+\textit{sim\_telarray} simulations. Note that if a muon has undergone significant scattering while traveling through the atmosphere, the impact position on the ground is no longer a meaningful parameter to describe its trajectory. For this reason, the comparison shown in Figure~\ref{fig:toycompare} is done with a subset of the full simulations with events selected for modest scattering. As expected, the agreement is very good. A further check utilizing the entire set of simulations is described below.

Once the properties of the photoelectrons produced by a muon and captured by the camera are known, the next step is to determine if they would activate the camera trigger. For this we base our trigger definition on the one used by FlashCam, the camera installed in the largest H.E.S.S. telescope since October 2019~\cite{fc-trigger}. The criterion is passed when an image has a group of nine neighboring pixels with a total of more than 68 photoelectrons. For each combination of muon starting height, initial energy and telescope distance from muon position, the corresponding muon image in the telescope is generated from the Cherenkov light distribution in the SMM and tested against this criterion. This allows for a trigger decision dependent on those three parameters only, which we will refer to as the simplified trigger criterion. 

To verify that this is an accurate description of the real trigger conditions we once again use the small sample of CORSIKA+\textit{sim\_telarray} simulations. For this comparison, we include muon events undergoing significant scattering, a process which, together with bremsstrahlung, is ignored by the SMM (see Table~\ref{tab:compare}). To do this, we compare, for different muon starting heights and as a function of energy and impact distance, the total number of muons arriving at the ground in the CORSIKA sample with the number of muons detected by the telescope after processing the sample with \textit{sim\_telarray}. Integrating the ratio of these numbers radially results in an effective area for a given energy and starting height. Similarly, for the same set of input CORSIKA events, we use their energy, production height and angular direction to compute our simplified trigger criterion. We then compute the muon effective area in the same way. These quantities are shown in the bottom panel of Figure~\ref{fig:toycompare}. This comparison includes muon events that undergo scattering and bremsstrahlung, confirming that the simplified trigger criterion is accurate to well within 10\% across all energies.

\begin{figure*}[b!]
    \centering
    \includegraphics[width=1.45\columnwidth]{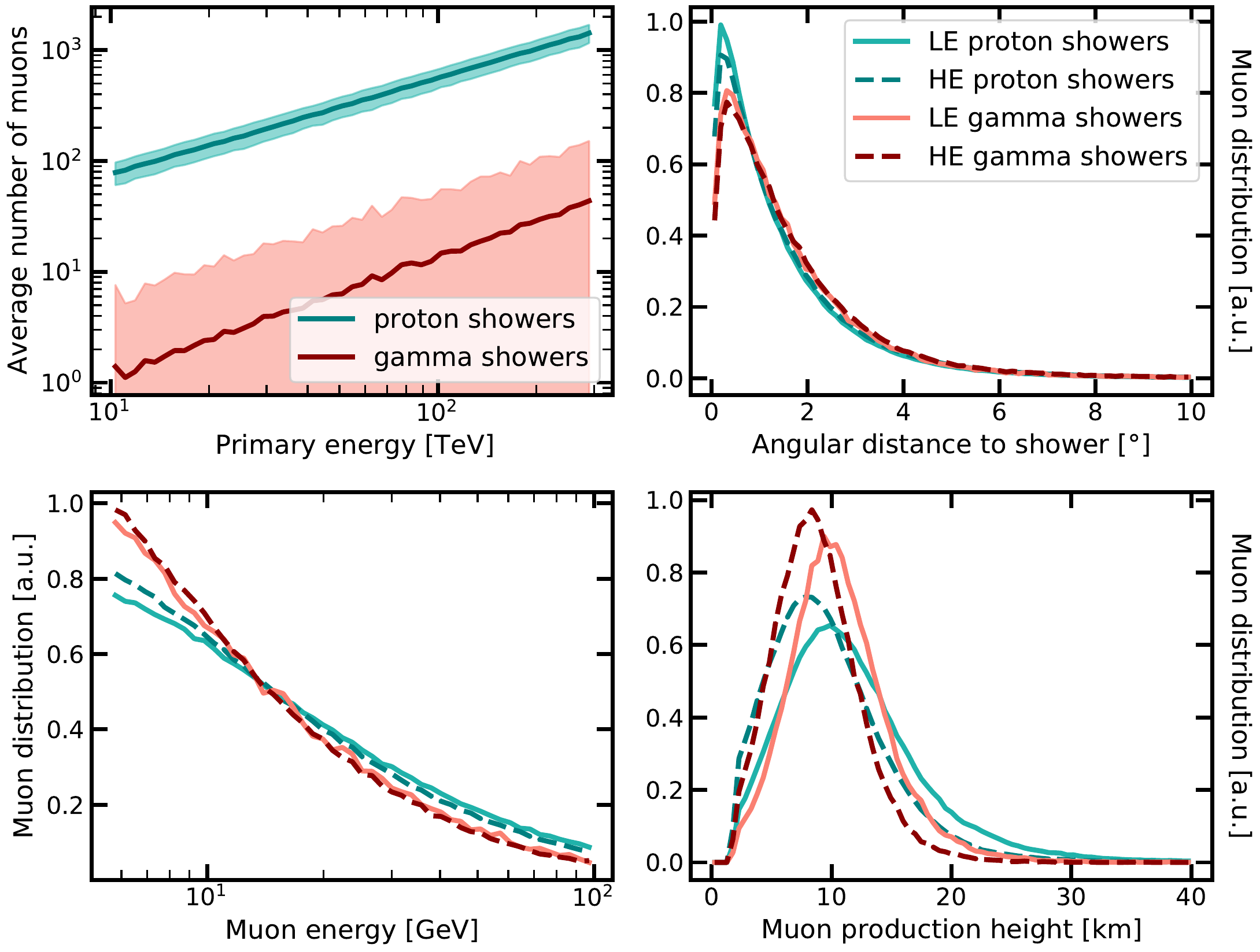}
    \caption{Characteristics of muons produced in proton- and gamma-ray-initiated showers. The top left panel shows the average number of muons present as a function of primary energy, with the shaded area representing the standard deviation. The remaining panels show the distributions of angular distance to the shower, energy and production height for muons produced by showers in two energy ranges, normalised for equal area. Solid lines correspond to ``low-energy''~(LE, 10-20 TeV) showers, while dashed lines represent muons in``high-energy''~(HE, 100-120 TeV) showers.}
    \label{fig:showermuons}
\end{figure*}

\newpage
The distribution of arrival times of the Cherenkov photons produced by the muon is straightforward to compute with the setup described above. Defining the instant in which a muon produced at height $h_{\mu}$ arrives at the ground as $t_{\mu}$, we can compute the relative delay in arrival time of a photon produced by the muon at height $h$, $t_{\gamma}(h)$ as
\begin{equation}
    \Delta t(h) = t_{\gamma}(h) - t_{\mu} = \frac{h\cdot c}{\cos \theta_c(h)} - t_{\mu},
\end{equation}
where $\theta_c(h)$ is the Cherenkov angle. Using Equation~\ref{eq:impact} we compute $\Delta t(R)$, that is, the relative photon arrival time as a function of the ground distance to the muon impact point. 

\begin{figure}
  \includegraphics[width=0.45\textwidth]{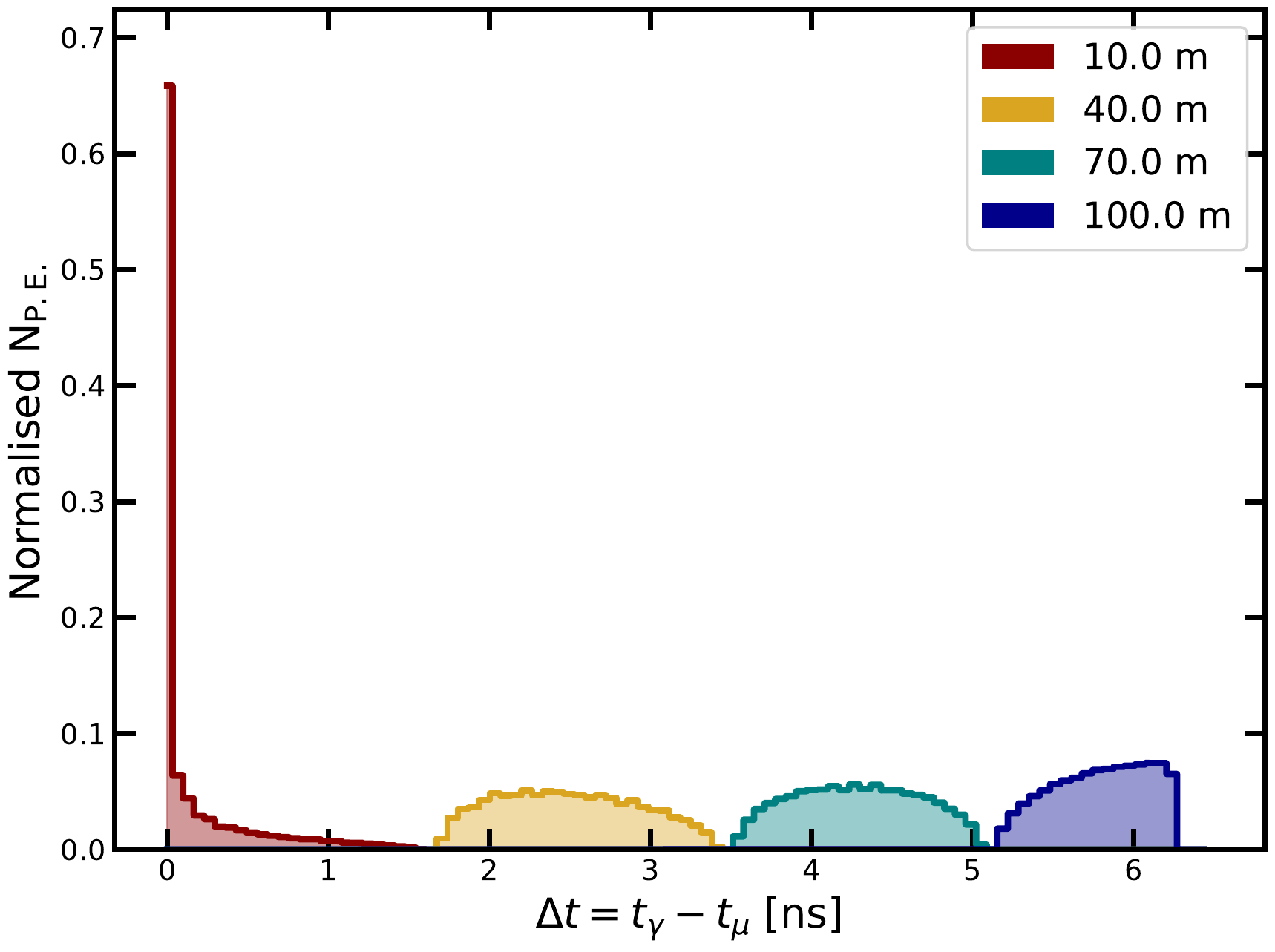}
\caption{Distribution of the arrival time of Cherenkov photons for a muon of energy 20~GeV and starting height 11~km as seen from several impact distances. The quantity in the y-axis is normalised for equal area at each impact distance.}
\label{fig:timedistr}       
\end{figure}

Now, combining this quantity with Equation~\ref{eq:pe} allows us to compute the number of photons that arrive to the dish of a telescope $\mathrm{N_{P.E.}}$ for each value of $\Delta t(R)$. An example of such distributions for a muon of energy 20~GeV can be seen in Figure~\ref{fig:timedistr}. Note that the distribution for each telescope location is normalised for equal area.

\section{Detectable muons in showers}
\label{sec:showers}
Considering shower development only down to the Cherenkov threshold for muons allows extremely rapid simulations with CORSIKA, even up to primary energies of many hundreds of TeV. For both gamma-ray- and proton-initiated showers, over $10^7$ showers were produced with energies between $10^1$ and $10^{2.5}$~TeV, distributed as $\propto E^{-1}$ to ensure enough events at the highest energies. For all showers, the primary particle initial direction was $\theta_z=20$\degree. From these showers, the production height, energy, ground level direction and impact point were extracted for all the muons present. Figure~\ref{fig:showermuons} summarises the basic properties of these muons.

The very high statistics allows us to probe the characteristics of the very rare most muon-poor proton showers, which form the irreducible background of the muon-tagging approach. We define a muon as \textit{detectable} if they fulfill three separate conditions, with the first being the simplified trigger criterion described in Section~\ref{sec:muons}. The remaining conditions refer to the muon trajectory and are described below. We assume for simplicity that the telescope is always pointed towards the shower axis. The second condition is then given by the fact that the angular distance between the muon trajectory and the axis must be smaller than half of the telescope field of view (FoV), taken here to be that of the central H.E.S.S. telescope, 3.5$\degree$.

\begin{figure}[]
    \centering
    \includegraphics[width=0.45\textwidth]{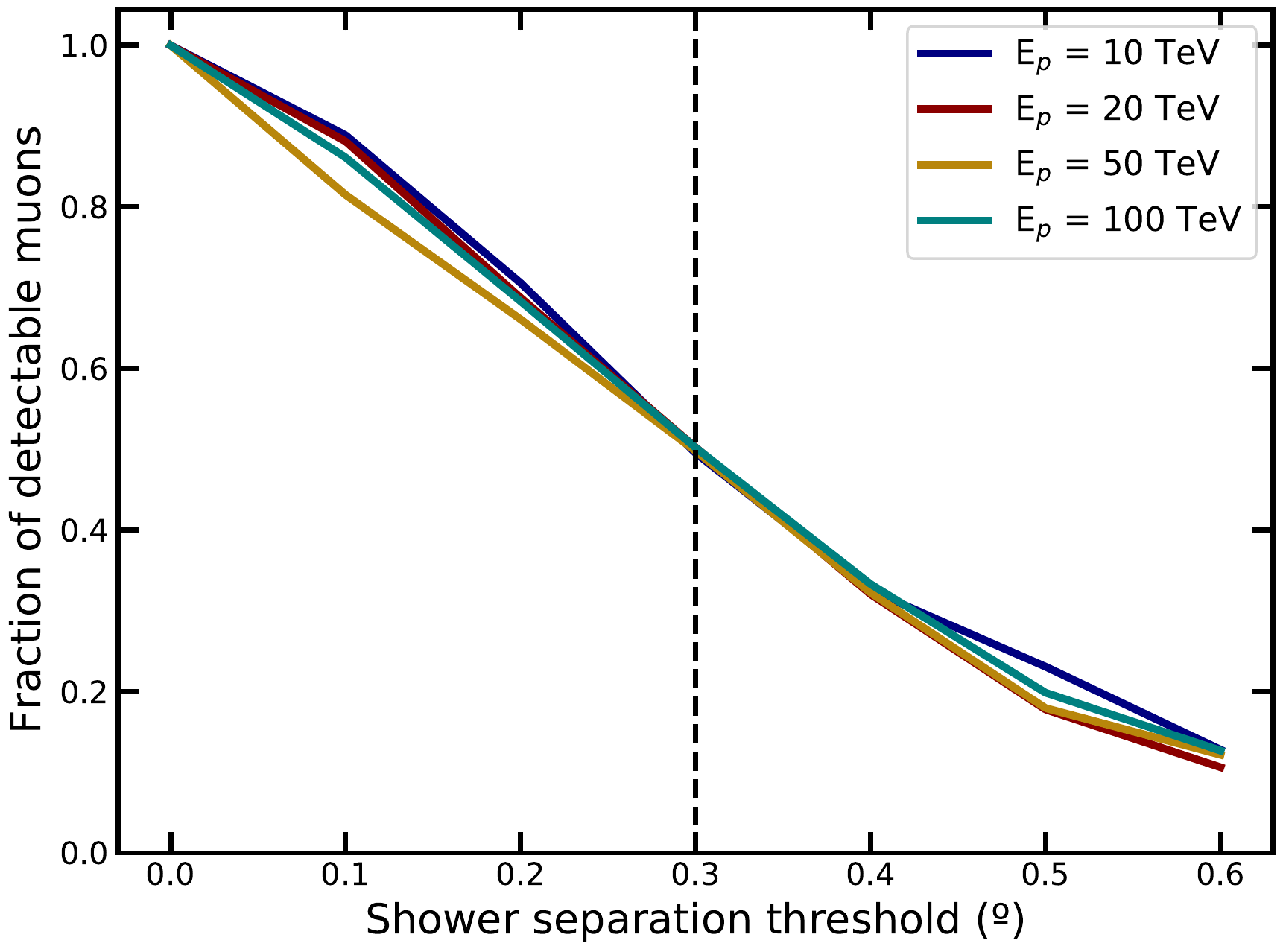}
    \caption{Fraction of surviving muons as a function of the minimum shower separation threshold for several energies. Both the trigger and FoV criteria have already been applied. The dashed line indicates the chosen reference value of 0.3$\degree$ which translates to roughly 50\% survival due to this criterion.}
    \label{fig:showerdistance}
\end{figure}

Beyond simply being able to detect light from muons, in order to use their presence as a background rejection criterion, it is crucial to accurately identify them as such. Muon identification can be carried out via different techniques, which exploit the properties of the muon Cherenkov signal described in Section~\ref{sec:muons}. A thorough study of the different muon identification techniques and their performance is beyond the scope of this paper (see~\cite{alisonmuons,muon-identification-hough,muon-identification-cnn} for some recent efforts). Current background rejection methods implemented in IACTs rely on the properties of the time-integrated Cherenkov light images. In this case, a complete overlap in the camera image between the main shower component and the light coming from muons makes this identification very difficult. In order to take this into account, and also to explore the effect of different muon identification efficiencies, we impose a third and final condition for detectability: a requirement on the minimum angular distance between the trajectories of the muon and the primary particle. We choose a reference value for this distance of 0.3$\degree$ which corresponds to the survival of, as can be seen in Figure~\ref{fig:showerdistance} around half of the muons that pass the first and second detectability conditions described above. 

Note that possible muon identification techniques that exploit the arriving times of Cherenkov photons would not be limited by image overlap. Additionally, even techniques that are based in the integrated image can also strongly be affected by the shape of the separable muon component rather than its size in pixels. Hence this requirement on the minimum angular distance is only a proxy of the fact that only a fraction of the detected muons will realistically be tagged as such. In any case, as we show later in Section~\ref{sec:results}, the background rejection power attainable with the muon tagging strategy would be competitive up to a muon identification efficiency of a few percent.

\begin{figure*}
\subfloat{%
\begin{overpic}
[width=0.96\columnwidth]{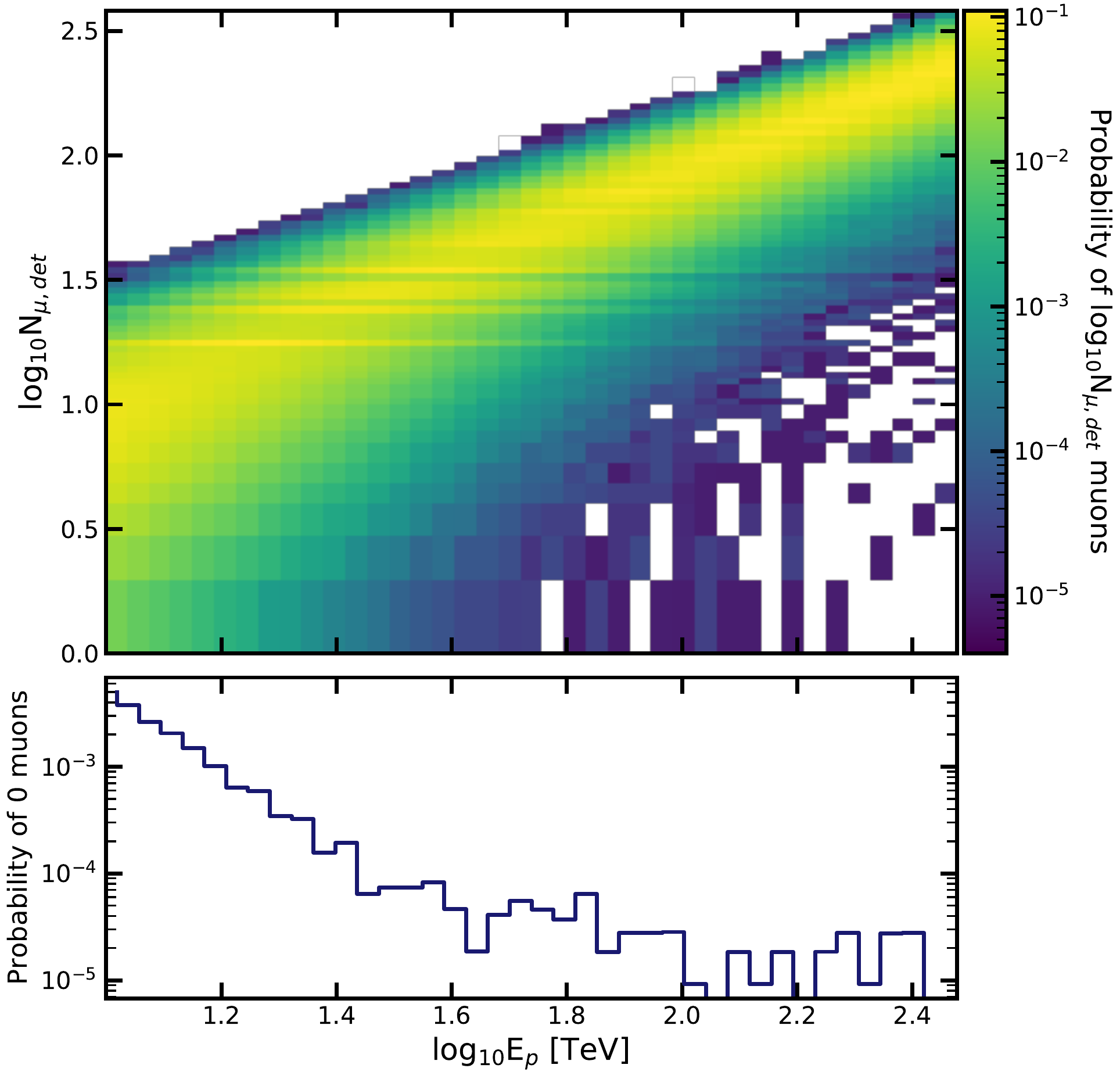}
\put(14,89){\Large p}
\end{overpic}
  } \hfill
\subfloat{%
  \begin{overpic}[width=0.96\columnwidth]{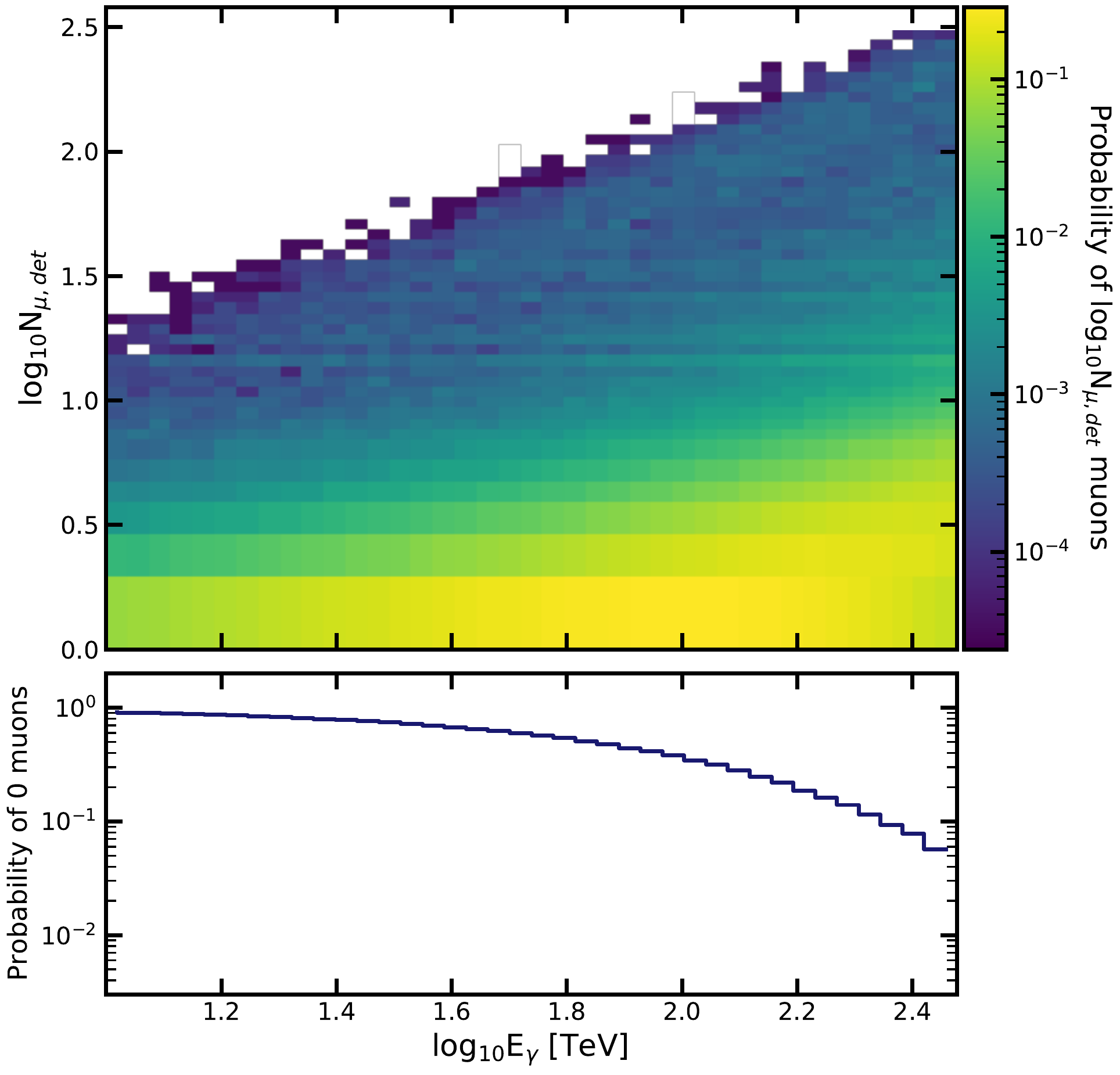}
  \put(14,89){\Large $\gamma$}
\end{overpic}
  }
\caption{\textit{Top panel}: Logarithm of the number of detectable muons in showers vs primary energy. The distribution is normalised at each primary energy bin. The type of particle initiating the showers is indicated on the top left corner. \textit{Bottom panel}: Probability of 0 muons in a shower vs primary energy. normalised together with the upper panel at each energy.}
\label{fig:2dhists}       
\end{figure*}

\begin{figure*}
\subfloat{%
\begin{overpic}
[width=0.45\textwidth]{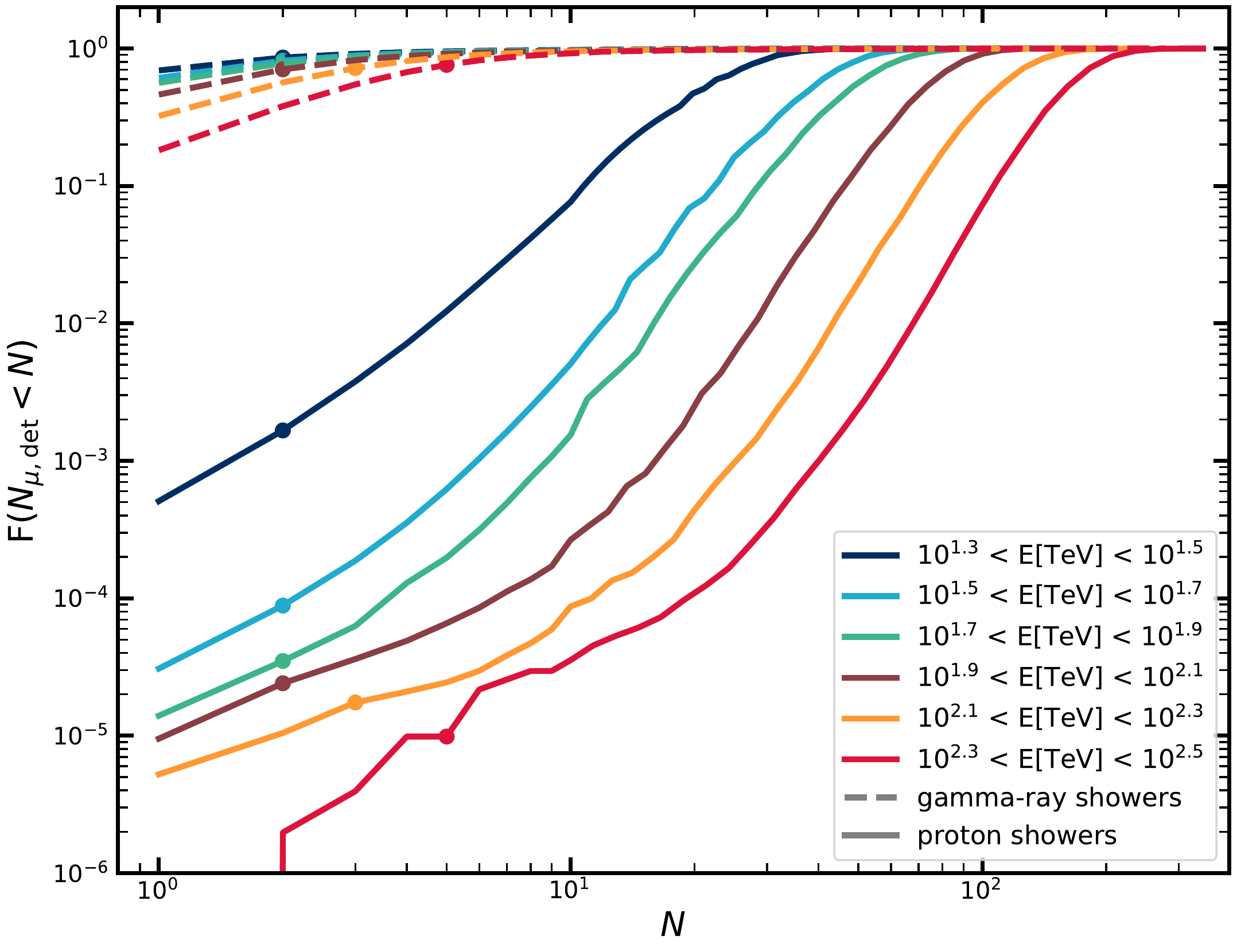}
\put(77,37){D$_T$=28~m}

\end{overpic}
 } \hfill
\subfloat{%
   \begin{overpic}
[width=0.45\textwidth]{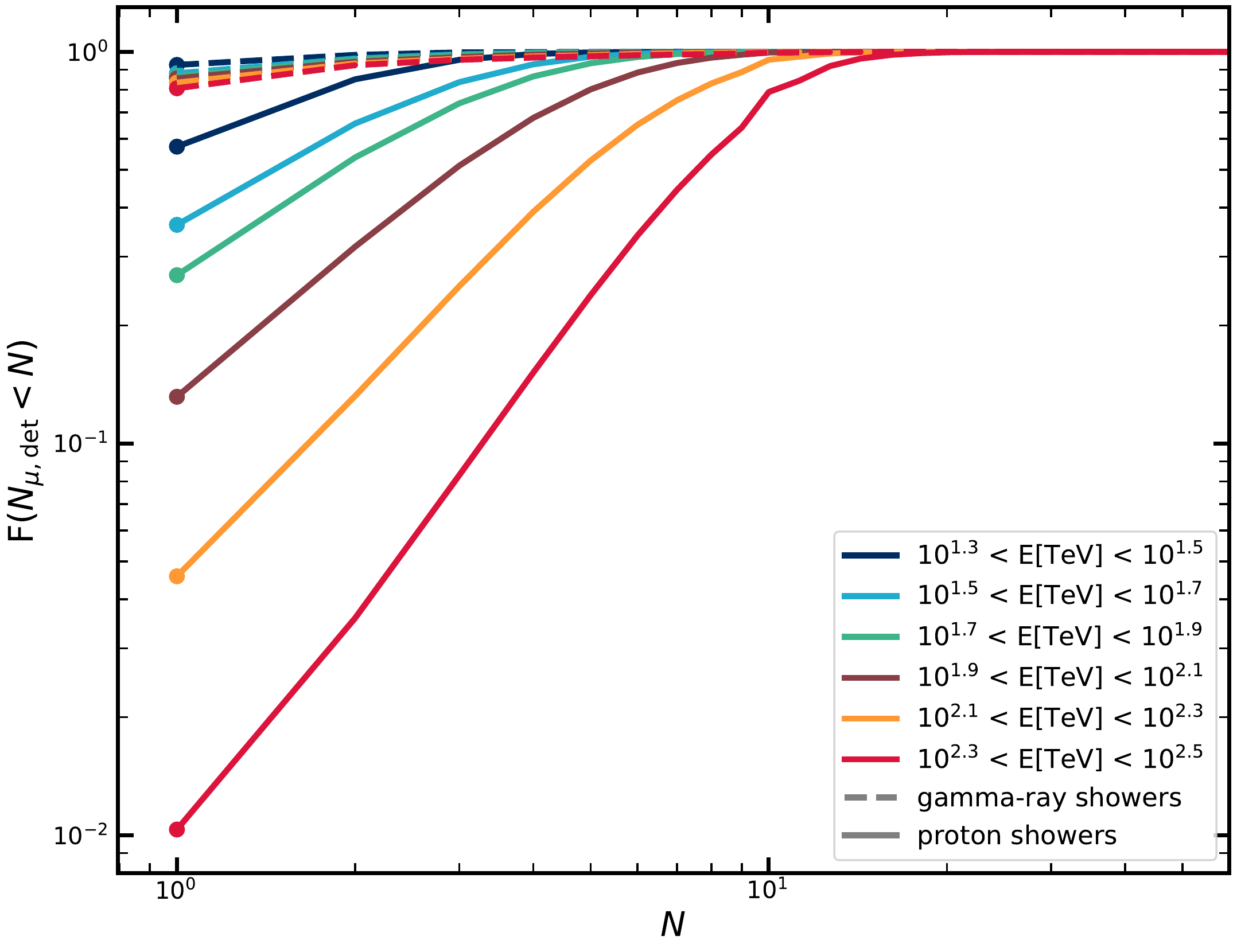}
\put(77,37){D$_T$=12~m}

\end{overpic}
  }
\caption{ Cumulative distribution of the number of detectable muons in proton-initiated showers (solid lines) and gamma-ray-initiated showers (dashed lines) for several primary energy ranges. A round marker is placed at the x-position defined by a 70\% signal efficiency. The corresponding telescope size is indicated with D$_T$ in each panel.}
\label{fig:cumulative}       
\end{figure*}
\section{Results}
\label{sec:results}
Figure~\ref{fig:2dhists} shows the logarithmic distribution of the number of muons classified as detectable, $\log_{10}N_{\mu, \mathrm{det}}$, for showers with both protons and gamma rays as the primary respectively. The energy of the primary particle energies ranges between 10 and 250 TeV. The differences between both distributions are striking, revealing that a large fraction of the initially high (see Figure~\ref{fig:showermuons}) number of muons in proton showers can be in principle detectable by a large telescope. Conversely, for gamma-ray showers, the distribution is shifted towards a much lower number of detectable muons. The bottom panel of both figures shows the probability that a shower does not contain any detectable muons.
Above 100~TeV this number is of the order of $10^{-5}$ for proton showers, a number which assumes a 50\% muon identification efficiency as described in Section~\ref{sec:showers}. This number represents the irreducible background of this approach, and is orders of magnitude below the current rejection power reached at those energies~\cite{DanStefan,stefanBDT,dan-HE}, even assuming  a less effective muon identification strategy.

Let us refer to the probability distributions of detectable muons shown in Figure~\ref{fig:2dhists} as $f(N_{\mu, \mathrm{det}} | E_{\mathrm{pri}})$, where $E_{\mathrm{pri}}$ is the energy of the primary particle. From this quantity, we can compute the fraction of showers expected to contain less then a certain number $N$ of detectable muons as:
\begin{equation}
    \mathrm{F}(N_{\mu, \mathrm{det}} < N | E_{\mathrm{pri}} ) = \int_{0}^{N} f(N_{\mu, \mathrm{det}} | E_{\mathrm{pri}})dN_{\mu, \mathrm{det}}
\end{equation}

Note that this corresponds to an integral as a function of the y-axis of each panel in Figure~\ref{fig:2dhists}. We show this quantity corresponding to several ranges of primary particle energies for both gamma-ray and proton showers in the left panel of Figure~\ref{fig:cumulative}. Beyond 30 TeV substantial separation power would clearly result from effective identification of single muons within shower images. For energies higher than 80 TeV the separation power could reach $10^{-5}$, well beyond that achieved so far for IACT arrays.

\subsection{Smaller telescopes}
The large collection area of the telescope is critical to the number of muons whose light is collected by the dish. The simplified model results shown in Figures~\ref{fig:2dhists} and the left panel of Figure~\ref{fig:cumulative} correspond to a telescope with a diameter of 28~m. Repeating the same process for a smaller telescope, with a diameter of 12~m yields very different results. As can be seen in the right panel of Figure~\ref{fig:cumulative}, the background rejection power achievable with the muon-tagging approach worsens dramatically by several orders of magnitude. This is due to the fact that large dishes translate to a larger light collection area. Hence more light can be collected by the PMTs, allowing the detection of a higher number of muons, which are faint emitters compared to showers.

\section{Discussion}
\label{sec:discussion}
It is clear from Figure~\ref{fig:cumulative} that very significant potential exists for improving the background rejection of IACT arrays containing at least one large telescope at the highest energies provided that individual muon arcs can be identified with reasonable efficiency in the presence of bright shower images. A very large fraction of the muons initially present in the shower can in principle be detected by the telescope, thanks in part to the fact that the typical angular displacement from the shower is small enough to remain inside the FoV, as can be seen in Figure~\ref{fig:showermuons}. In order to exploit this potential, effective muon identification is required. There are several promising strategies, which use different characteristics of the muon Cherenkov light. As shown in Figure~\ref{fig:timedistr}, light produced by muons arrives to the telescope in short, concentrated bursts, unlike the shower photons, which span a range of tens of nanoseconds~\cite{timedistr}. This difference could be exploited to identify muons if the time distribution of the photons, rather than only the time-integrated images, is available~\cite{timing}. 

On the other hand, for time-integrated images, it has become very common and effective to employ template fitting approaches~\cite{impact,model++} during the reconstruction. However, very energetic events produce very bright and large shower images, which are often not entirely contained in the camera field of view. The very bright pixels from the shower image likely dominate the fit and obscure the presence of dim, constant surface-brightness muon images. Simply masking out the main shower image and analyzing the residual emission will likely result in an increased rate of muon identification, provided that the image cleaning procedure is not excessively harsh. More sophisticated approaches, making use of pixel-based deep learning techniques have already shown some promising results~\cite{muon-identification-cnn}. The recent development of dedicated packages such as CTLearn~\cite{CTLearn} or GammaLearn~\cite{GammaLearn} will likely facilitate the application of such techniques to the muon identification issue.

Large Cherenkov telescopes are widely seen as a useful asset for the low-energy range of the very-high energy spectrum due to their reduced threshold, but of limited use when considering the highest energies. However, as can be seen by comparing both panels of Figure~\ref{fig:cumulative}, this same reduced threshold is key when it comes to detecting the faint light coming from muons, which are overwhelmingly more common in proton-induced showers (see Figure~\ref{fig:showermuons}). Exploiting this potential of large telescopes through efficient muon identification algorithms could provide significant improvements in background rejection above several tens of TeVs, and in turn, improve the instrument sensitivity at the highest energies.

Another option is, of course, to build ground level muon detectors. However, these are not planned for either the existing or upcoming IACT arrays. Additionally, an improved muon detection technique could be applied retroactively to the entire data archive from an observatory, in the case of existing IACT arrays. This would effectively, and without increased observation time or hardware improvements, increase the detection capability at higher energies.

The results in Section~\ref{sec:results} were produced for showers arriving from a distance of 20\degree~from zenith. We also produced smaller samples of showers from 0\degree~and 40\degree~to explore the effect of the zenith angle on the result.  With increasing zenith angle, the number of detected muons goes up slightly, since the distance out to which muons are able to trigger goes up (see Equation~\ref{eq:impact}).

As can be seen in Figure~\ref{fig:2dhists}, muons are often produced, albeit in low numbers, by the highest energy gamma-ray showers. This is because at those energies, the number of interactions is so large that rare processes, such as muon pair production, become relevant. This indicates that perhaps the most useful separation parameter in terms of gamma-ray efficiency might not solely be the identification of individual muons, but rather a measure of the muonic content of an event.

The hadronic background is in fact not solely made up by protons, but also heavier nuclei~\cite{crays}. The muon content in these showers is higher, which translates into a higher number of expected detectable muons. In a sense, the proton case is the worst-case scenario, being the hadronic background species most likely to initiate muon-poor showers. However, protons are most likely to masquerade as a gamma ray in traditional background rejection approaches~\cite{2007APh....28...72M}, and therefore the room for improvement is greater.

\section*{Declarations}

\textbf{Funding} This research was supported by the Max Planck Society and ETH Zurich. AMWM is supported by the Deutsche Forschungsgemeinschaft (DFG, German Research Foundation) – Project Number 452934793, MI 2787/1-1.\\
\textbf{Conflicts of interest} Not applicable.\\
\textbf{Data Availability Statement} Data sharing not applicable to this article as no datasets were generated or analysed during the current study.\\
\textbf{Code availability }Not applicable.\\

\begin{acknowledgements}
The authors would like to thank the H.E.S.S. Collaboration for allowing the use of H.E.S.S. simulations in this publication, as well as providing useful discussions and input to the paper.
This work made use of \texttt{numpy}~\cite{numpy}, \texttt{scipy}~\cite{scipy}, \texttt{pandas}~\cite{pandas} and \texttt{matplotlib}~\cite{matplotlib}.
\end{acknowledgements}

\bibliographystyle{number_cite}       

\bibliography{refs}   

\end{document}